# Characterizing Coronal Mass Ejections in Solar Cycle Analysis

*[1,2]Ryan Manuel D. Guido

[1]Department of Earth and Space Sciences, Rizal Technological University, Philippines

[2]College of Graduate Studies and Teacher Education Research, Philippine Normal University, Philippines

**ABSTRACT**

The Sun is the major source of heat and light in our solar system. The solar cycle is an 11-year cycle of solar activity that can be determined by the rise and fall in the numbers and surface area of sunspots. Solar activity is associated with several factors including radio flux, solar irradiance, magnetic field, solar flares, coronal mass ejections (CMEs), and solar cycles. This study attempts to determine the Sun's activity specifically for the coronal mass ejection, its trend during solar cycle 23, and the apparent differences. A time series analysis was used to measure the CME data for larger cases and to see the apparent difference and trends of the CMEs. The result shows that a decreasing trend of coronal mass ejection from the year 1996 to 2016. It is therefore concluded that the coronal mass ejection data are normally distributed while coronal mass ejections are distributed normally and curved as fluctuation was found in the intensity of the disturbed storm time index as the number of great geomagnetic storms undeniably increaed in the ascending and the descending phases of the cycle. This means that even though the Sun has cycles and trends in its inherent characteristic, the Sun still possesses getting more dynamic through time which showcases that through the limited parameters involved in this study.

## 1. Introduction

The Sun is the major source of heat and light in our solar system. The solar cycle is an 11-year cycle of solar activity that can be determined by the rise and fall in the numbers and surface area of sunspots. Solar activity is associated with several factors including radio flux, solar irradiance, magnetic field, solar flares, coronal mass ejections, and solar cycles.

The geomagnetic activity relates to solar processes was confirmed by mid-nineteenth century: Solar flare correlation with active, bright auroras and geomagnetic disturbances was found by Carrington in 1860, and long-term observations showed 11-year variability both in sunspot numbers and occurrence frequency of magnetic disturbances and auroras (Pulkkien, 2007).

Several studies showed that multiple interacting magnetic clouds, as a result of the release of successive CMEs, were involved in a significant number of intense storms (Wang et al., 2003; Xie et al., 2006; Yermolaev & Yermolaev, 2008). Since geomagnetic storms can affect human lives, this study will determine the implications of geomagnetic activity on Earth's atmosphere. The 10 strongest CMEs during the solar cycle 24, and solar energetic particle events with intensity $\geq 10$ pfu in the $> 10$ MeV energy channel are significant in causing space weather effects and are commonly referred to as large SEP events are considered in this study. Major geomagnetic storms are those with geomagnetic indices (*Dst*) $< -100$ nT and are mostly caused by high-energy CMEs heading toward Earth (Gopalswamy, 2007). The goal of this research is to find the behaviors of the 10 strongest GMS of solar cycle 24, as currently recorded.

## 2. Solar Activity

The Sun undergoes various active processes that can be broadly regarded as solar activity. The presence of magnetic activity, including stellar flares, is considered as a common typical feature of Sun-like stars (Maehara, et al. 2012). Although a direct projection of the energy and occurrence frequency of superflares on Sun-like stars (Shibata, et al, 2013) does not agree with solar data

(Aulanier, et al, 2013) and terrestrial proxy, the existence of solar/stellar activity is clear (Usoskin, 2017).

The solar activities main feature is the quasi-periodicity with a period of about 11 years that varies in both amplitude and duration, known as the Schwabe cycle. Schwabe cycle is the most prominent variability in the sunspot-number series. It is renowned now as a vital feature of solar activity originating from the solar-dynamo process. It is blatant in many other parameters including solar, heliospheric, geomagnetic, space weather, climate, and others. Many indices are used to quantify diverse aspects of the variable solar activity. Quantitative indices comprise direct and indirect, they can be physical or synthetic. The longest available index of solar activity is the sunspot number, which is a synthetic index and is useful for the quantitative representation of overall solar activity outside the grand minimum. Furthermore, it showed that solar activity contains essential chaotic/stochastic components, which lead to irregular variations and make the prediction of solar activity for a timescale exceeding one solar cycle impossible (Usoskin, 2017).

According to the study of de Toma, et al (2013) solar cycle 23 was unlikely from the two previous cycles in many aspects. It showed that magnetic activity dropped during the maximum of cycle 23, also it slowed and terminated in a long and deep minimum characterized by a significant lack of sunspot activity and weak polar magnetic fields.

It is also evident in the study conducted by Georgieva and Kirov (2006) that there are common deviations in surface air temperature that follow the variations in solar activity showing the solar influences on climate, in the last few decades solar activity has persisted more or less constant while temperature has continued increasing which is a strong argument in favor of anthropogenic influences on climate.

**3. Coronal Mass Ejection**

Coronal Mass Ejections (CMEs) hurl huge masses of energized gas out into space. These occur most often during the period of maximum solar activity and cause a phenomenon called space weather. These are the outburst of solar energetic particle events, as a result of acceleration and heating of solar plasma during solar flares. Geomagnetic Storms (GMS) are caused by the interactions by materials ejected from the Sun, specifically, Solar Energetic Particle (SEP) events that lead to the disturbances in the Earth's magnetic field (Guido, 2016). Increases in the number of solar flares and CMEs raise the probability that complex instruments in space will be impaired by these accelerated energetic particle events The SEP can also threaten the health of both astronauts in space and airline travelers in high-altitude, polar routes (Pulkkien, 2007).

The interaction of the geomagnetic field with the magnetic field carried within CMEs and the surrounding background magnetized by the modulated solar wind. If the speeds range from 400-2,500 km/sec, it takes some 1-4 days for CMEs to propagate from the Sun to the Earth, with typical transmit time of 2-3 days. Correlations between the strength of CMEs, and the magnitude of their impact in geospace continue to be studied, both observationally and in numerical analyses (Newell, et al., 2007; Schrijver, 2009; Andreeova, et al., 2011).

It also showed that the geomagnetic disturbances serve different space weather hazards, ranging from satellite system to ground facilities, such as induction current, drastic variation of radiation belt particle flux, heating and expansion of polar upper atmosphere, and development of ionospheric storms in which all these phenomena have been a subject under intensive space weather research and are worth being predicted accurately in practical space weather forecast Miyake & Nagatsuma (2012).

A geomagnetic activity that is measured by *Dst* is principally driven by the plasma and magnetic field conditions in the solar wind that encounters the Earth (Tsurutani & Gonzales, 1997; O'Brien & McPherron, 2000). Indices such as *Dst* are used to assess magnetic storms severity. Thus, a *Dst* index below -50 nT is indicative of moderate disturbance, which turns to intense when -100 nT thresholds is passed (Gonzales et al., 1994) and super intense or extreme if *Dst* reaches less than -250 nT (Echer et al., 2008).

Miyake & Nagatsuma (2012) revealed that the geomagnetic disturbances are more difficult to be predicted than quiet intervals, suggesting that the simple correlation method of solar wind measurement at separated solar longitude is not enough to accurately predicting geomagnetic disturbances, even though the correlation seems generally high.

The solar cycle distribution of major geomagnetic storms ($Dst \leq -100$ nT), including intense storms at the level of $-200$ nT $< Dst \leq -100$ nT, great storms at $-300$ nT $< Dst \leq -200$ nT, and super storms at $Dst \leq -300$ nT, which occurred during the period of 1957-2006, based on $Dst$ indices and smoothed monthly sunspot numbers. It also shows that the majority (82%) of the geomagnetic storms at the level of $Dst \leq -100$ nT that occurred in the study period were intense geomagnetic storms, with most are ranked as great storms and almost half of it is super storms (Le et al., 2013).

Geomagnetic storms are largely associated with CMEs from the sun. CMEs faster than $-500$ Km/s eventually drive shock waves which normally strike the earth's magnetosphere in 24 to 36 hours after the even on set on the sun. 47 geomagnetic storm with minimum Dst $\leq -100$nT.

The Carrington event of 1859, the March 1989 storm responsible for the Quebec power outrage or the October 2003 storm threatening the electrical grid in South Africa and Sweden cannot be missed in the short list of historical records of extreme geomagnetic storms due to their consequences for society. All these three storms had consequences in infrastructures, and all of them were super intense as seen by $Dst$ index (Cid et al., 2014). Carrington event was estimated by Lakhina et al. (2005) as -1760 nT, approximately three times more intense than the Quebec storm.

It is concluded by Selbergleit (2015) that the localized Earth potentials can produce significant effects on power systems and pipelines. The corrosion is increased by the electric currents that spread through the ground during magnetic storms and substorms; the difference in potential with the ground (PV) can become positive by several volts, resulting in electron leakage. It is also observed the PV values were always negatives, but during some periods, they were greater than -850 mV.

It is showed in the study of Mansilla and de Artigas (2010) that the SEP event considered as one of the 4 major events of the solar cycle 23, which has a maximum proton flux of 24,000 pfu. During strong SEP events (and intense geomagnetic storms), the solar protons and the auroral electrons possibly have sufficient energy as to penetrate to the height covered by the meteorological balloons and so, these charged particles could give rise to weak and disperse changes in temperature. Statistically significant increase in temperature is not observed in the study. Further, studies in different latitudinal sectors are necessary in order to determine whether there are temperature increases depending on the intensity of the geomagnetic storm.

Studies about the role of solar variability found out that there is a strong correlation between points of surface air temperature. Implying that solar variability has been the dominant influence on northern hemisphere temperature trends since at least 1881. That there is an apparent correlation, and its implications for previous studies which have instead suggested that increasing atmospheric carbon dioxide has been the dominant influence (Soon, et al, 2015).

## 4. Materials and Methods

This study will give rise to the study of solar science that will showcase the exquisite role of the Sun through analyzing its behavior in terms of its coronal mass ejections outburst.

The data were gathered from Compact Astronomical Low Cost, Low Frequency Instrument for Spectroscopy and Transportable Observatory (CALLISTO) network: STEREO, LASCO, SDO/AIA, and Solar Heliospheric Observatory (SOHO), Space Weather Services of the Australian Government Bureau of Meteorology, Sunspot Index and Long-term Solar Observations (SILSO), Solar Physics, Marshall Space Flight Center, National Aeronautics and Space Administration (NASA), Space Weather and Prediction Center, National Oceanic and Atmospheric Administration (NOAA), to obtain accurate results of the data.

A time series analysis was used to determine the trend of the CME data for larger cases. Fourier analysis of the Disturbed Storm Time Index of the Solar Cycle 23 to determine the ascending and descending phase of the solar cycle and the relationship that could determine with the peak during this solar cycle, and Kolmogorov-Smirnov Test was used to determine the apparent difference of CME and its specific dispersal over time.

## 5. Results and Discussions

The result of the time series trend analysis of coronal mass ejections shows the increase and decrease since the start of data collection from 1996 to 2016.

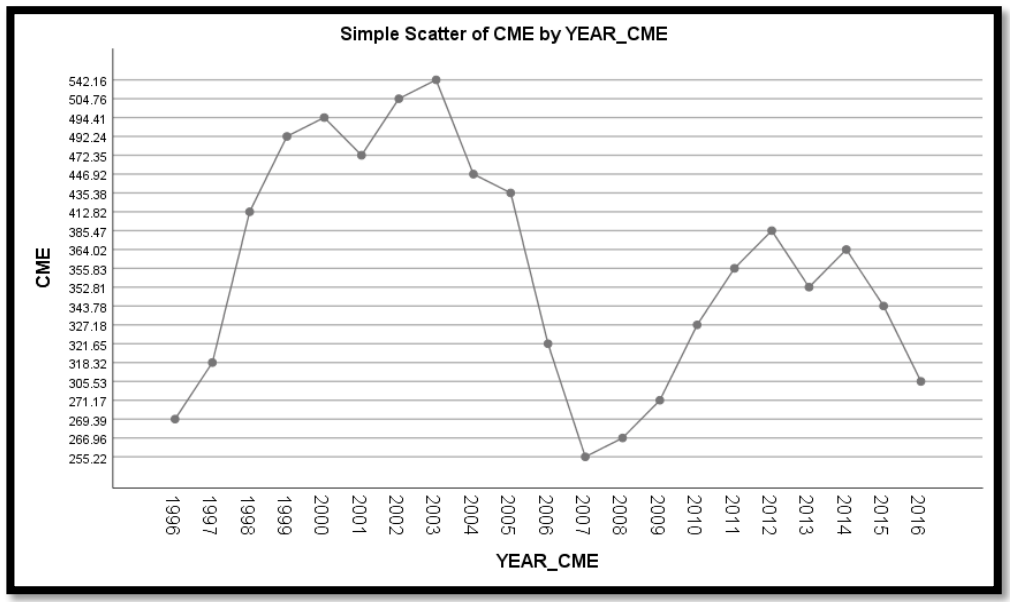

**Figure 1:** Time Series Trend Analysis of Coronal Mass Ejections

It further shows that the trend is decreasing. This basically reveals that the amount of coronal mass ejections being generated by the Sun is decreasing as time goes by which means that the sun produces lesser ejections. It is also noted that the rise of the CMEs was visible during 2003 and a deep decrease during 2007. Both peak and crest of the figure are within the solar cycle 23. Thus, interposing no probable reason, for now, the influence of the peak and crest of the CME in the same solar cycle.

Miyake & Nagatsuma (2012) showed that the geomagnetic disturbances cause various space weather hazards, ranging from satellite system to ground facilities, such as induction current, drastic variation of radiation belt particle flux, heating and expansion of polar upper atmosphere, and development of ionospheric storms in which all these phenomena have been a subject under intensive space weather research and are worth being predicted accurately in practical space weather forecast.

Table 1 represents the great geomagnetic storms during the solar cycle 23. It shows the events or dates where the great geomagnetic storms occur. The solar cycle 23 started from May 1996 up to December 2008. The table reveals that the first occurrence of a great geomagnetic storm on Earth occurred last May 4, 1998, with Dst -205 nT and the last occurred on August 24, 2005, with Dst -216 nT.

**Table 1:** The Disturbance Storm Time Index during the Major Geomagnetic Storms of Solar Cycle 23

| Geomagnetic Storm Events | Disturbance Storm Time Index Dst/nT |
|---|---|
| **5/4/1998** | -205 |
| **10/22/1999** | -237 |
| **4/7/2000** | -288 |
| **7/16/2000** | -301 |

| 8/12/2000 | -235 |
|---|---|
| 9/17/2000 | -201 |
| 3/31/2001 | -387 |
| 4/11/2001 | -271 |
| 11/6/2001 | -292 |
| 11/22/2001 | -221 |
| 10/30/2003 | -353 |
| 11/20/2003 | -422 |
| 11/7/2004 | -373 |
| 11/10/2004 | -289 |
| 5/15/2005 | -263 |
| 8/24/2005 | -216 |

From the table above it can be interpreted that there is a fluctuation on the amount of the disturbed storm time index, there are noticeably raised number of great geomagnetic storms during the ascending and descending phase of the solar cycle.

It also showed up during the events of 2000 and 2001, there are multiple entries relating great geomagnetic events as compared to other years because this was the time where the Sun has reached its peaked and NASA reported that February 2001 that the Sun's magnetic field had flipped. The flip means that the Sun's north pole, which had been in the northern hemisphere of the Sun flipped into the southern hemisphere. This normally happens during the peak of each solar cycle.

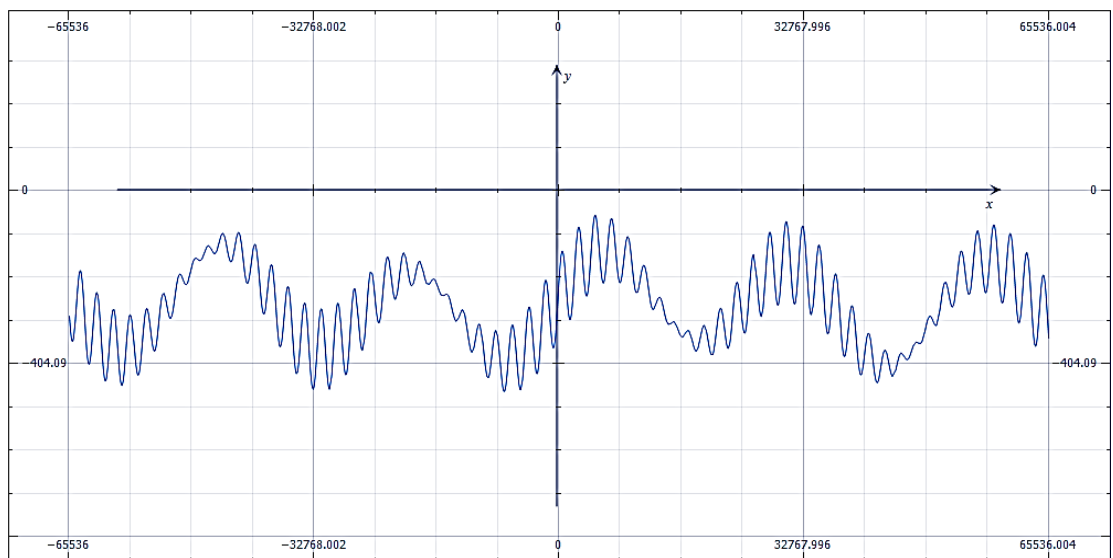

**Figure 2:** Fourier Analysis of the Disturbed Storm Time Index of the Solar Cycle 23

Figure 2 reveals the Fourier analysis of the Disturbed Storm Time Index of the Solar Cycle 23 to determine the ascending and descending phase of the solar cycle and the relationship that could determine with the peak during this solar cycle.

This solar cycle noticeably has raised a number of great geomagnetic storms in the ascending phase however, there are more peaks during the descending phase. The solar cycle 23 started in April 1996 and had its peak in early 2000 and 2001. The decline phase of this period extended from 2002 until the latter part of 2008, this shows that this is the longest declining phase of all solar cycles occurred. Solar cycle 23 lasted for about 13.5 years but is considered to be the weak solar cycle for the past 23 solar cycles.

This solar cycle minimum seems to have unusual properties that appear to be related to week solar polar magnetic fields and magnetic field activity during this solar cycle has been very weak with sunspot numbers reaching the lowest values in about 100 years (Hathaway, 2011; Hady, 2014).

**Table 2:** Test of Normality of the Coronal Mass Ejection Data

**Tests of Normality**

|  | Kolmogorov-Smirnov[a] | | | Shapiro-Wilk | | |
|---|---|---|---|---|---|---|
|  | Statistic | df | Sig. | Statistic | df | Sig. |
| CME0 | .170 | 13 | .200* | .892 | 13 | .105 |

*. This is a lower bound of the true significance.
a. Lilliefors Significance Correction

Table 2 shows the test of normality of the coronal mass ejection data from 1996 to 2016. The yearly coronal mass ejection data has sig .200 > α = .05. The values of Kolmogorov-Smirnov and Shapiro-Wilk test are .200 and .105, respectively, and is greater than .05, it implies that it is acceptable to assume that the weight distribution is normal. With this, this appears statistically unlikely, so there is sufficient evidence that it failed to reject the null hypothesis. This indicates that it failed to reject the null hypothesis and therefore conclude that the coronal mass ejection data are normally distributed.

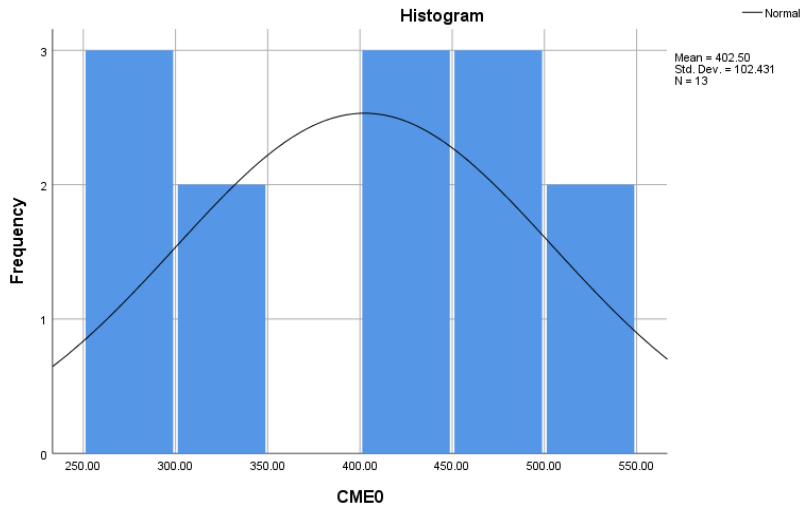

**Figure 3:** Histogram of the Coronal Mass Ejection Data

Figure 3 shows the diagram on the normal parameters of the frequency towards the coronal mass ejection. It shows that the coronal mass ejection is curved towards the center as distributed by the coronal mass ejection.

**6. Conclusion**

This study reveals that a decreasing trend of coronal mass ejection from the generated data from the year 1996 to 2016. This basically implies that the amount of coronal mass ejections being generated by the Sun is decreasing as time goes by which means that the sun produces lesser ejections. It is also noted that the rise of the CMEs was visible during 2003 and a deep decrease during 2007. Both peak and crest of the figure are within the solar cycle 23. The solar cycles have a larger number of great geomagnetic storms in the ascending phase, however, more peaks occur during the descending phase. It implies that it is acceptable to assume that the weight distribution is normal. With this, this appears statistically unlikely, so there is sufficient evidence that it failed to reject the null hypothesis.

The result shows that a decreasing trend of coronal mass ejection from the year 1996 to 2016. It is therefore concluded that the coronal mass ejection data are normally distributed while coronal mass ejections are distributed normally and curved. This means that even though the Sun has cycles and trends in its inherent characteristic, the Sun still possesses getting more dynamic through time which showcases that through the limited parameters involved in this study.